\documentclass[12pt]{article}

\usepackage[margin=1in]{geometry}
\usepackage{amsmath,amsthm,amssymb}
\usepackage{braket}

\usepackage{graphicx}

\usepackage{cite}

\usepackage{xparse}
\usepackage{cleveref}
\begin{document}

\begin{center}

\textbf{
Pole decomposition of BFKL eigenvalue at zero conformal spin and the real part of digamma function. }

\small
Mohammad Joubat$^{(a)}$, Claudelle Capasia Madjuogang Sandeu$^{(b)}$ 
   and Alex Prygarin$^{(b)}$  
\\ \nonumber
$^{(a)}$ Department of Mathematics, Ariel University, Ariel 40700, Israel\\ \nonumber
$^{(b)}$ Department of Physics, Ariel University, Ariel 40700, Israel 
\end{center}
\normalsize

\normalsize

\begin{abstract}
 We consider the powers of leading order eigenvalue of the Balitsky-Fadin-Kuraev-Lipatov~(BFKL) equation  at zero conformal spin. Using reflection identities of harmonic sums we demonstrate how involved generalized polygamma functions are introduced by pole separation of a rather  simple digamma function. This generates higher weight generalized polygamma functions at any given order of perturbative expansion. As a byproduct of our analysis we develop a general technique for calculating powers of the real part of digamma function in a pole separated form. 
\end{abstract}


\section{Introduction}

 Polygamma functions appear in solutions of integrable models 
 in High Energy Physics. The most famous example is the Balitsky-Fadin-Kuraev-Lipatov~(BFKL) equation~\cite{BFKL1,BFKL2,BFKL3,BFKL4}, which can be written as Schroedinger equation with eigenvalue expressed in terms of polygamma functions and their generalizations in the leading and the next-to-leading order~\cite{NLO1,NLO2,NLO3}. In particular, the scattering amplitude in the BFKL approach can be written in integral representation, where the integrand is an exponent of  real part of the digamma function in the leading order of the perturbation theory. Powers of real part of the digamma function were not studied before and there is no close expression for it in a general case. The main objective of the present study is to fill this gap deriving analytic expressions for the first few powers of real part of the digamma function, which can be further applied for analysing the eigenvalue of the BFKL equation at higher orders of perturbation theory.
 
 The paper is written as follows. In the first section
 we review the eigenvalue of the BFKL equation as well as  the  definition and general properties of the digamma function relevant for our analysis.  We show how the reflection identities of harmonic sums can be used for representing the powers of real part of the digamma function in pole decomposed form. We comment on complexity of the resulting functions emphasizing their analytic structure. In the second section we extend our analysis to the third and the fourth power of the real part of the  digamma function, and present some technical details of our   calculations. 
In the last section we summarize   the  results and  discuss possible applications  of the obtained expressions. 

\section{BFKL eigenvalue and the real part of digamma function }

The Balitsky-Fadin-Kuraev-Lipatov~(BFKL) equation~\cite{BFKL1,BFKL2,BFKL3,BFKL4} describes the energy dependence of the imaginary part of  scattering amplitude in Regge kinematics. The BFKL equation can be written in the form of the Schroedinger equation $H \psi =E \psi$, where $H$ is the Hamiltonian, which depends on transverse degrees of freedom, and $E$ is its eigenvalue depending on the coupling constant. Higher order of the perturbation theory, the higher powers of the coupling constant. The exact form of the BFKL eigenvalue is known only in the leading and next-to-leading order in QCD.  In $N=4$ super Yang-Mills theory the situation is much better, where in some limits it was recently calculated at higher loop orders (for recent review see \cite{Caron-Huot:2020grv} and references therein).

In the present paper we focus on the leading order of the BFKL eigenvalue 
\begin{eqnarray}
E \propto a  \left[  \psi \left(\frac{1}{2}+i \nu +\frac{n}{2}\right)+\psi \left(\frac{1}{2}-i \nu +\frac{n}{2}\right)- 2\psi \left(1\right)\right] = 2 a \; \Re \left[\psi \left(\frac{1}{2}+i \nu +\frac{n}{2}\right)- \psi \left(1\right)\right]
\end{eqnarray}
which is written in terms of the digamma function and two real variables, the continuous $\nu$ and   integer $n$~(conformal spin). The powers of the leading order  BFKL eigenvalue appearing in the higher loop perturbative expansion are the powers of the real part of the  digamma function $\psi(z)$. 
The digamma function is defined as logarithmic derivative of the Euler gamma function, i.e.
\begin{eqnarray}
\psi(z)=\frac{d \ln \Gamma(z)}{dz}=\frac{\Gamma'(z)}{\Gamma(z)}
\end{eqnarray}
Its integral and  series representation read
\begin{eqnarray}
\psi(z)=\int_0^\infty \left( \frac{e^{-t}}{t} -\frac{e^{-zt}}{1-e^{-t}} \right) dt
\end{eqnarray}  
and 
\begin{eqnarray}
\psi(z)= \sum_{k=1}^{\infty}\frac{z-1}{k(k+z-1)}+\gamma
\end{eqnarray}
respectively. Polygamma functions $\psi^{(m)}(z)$ are defined as derivatives of the digamma function 
\begin{eqnarray}
\psi^{(m)} (z)=\frac{d^m }{d z^m} \psi(z).
\end{eqnarray}

The  BFKL eigenvalue at the next-to-leading order \cite{NLO1,NLO2,NLO3} includes generalizations of the polygamma functions such as 
\begin{eqnarray}\label{functionsNLO}
  \sum_{k=0}^{\infty} \frac{\beta'(k+1)}{k+z},\;\; \;\sum_{k=0}^{\infty} \frac{(-1)^k\psi'(k+1)}{k+z}, \;\;\;\sum_{k=0}^{\infty} \frac{(-1)^k\left( \psi(k+1)-\psi(1)\right)}{(k+z)^2},
\end{eqnarray}
where 
\begin{eqnarray}
\beta'(z) =\frac{1}{4} \left[ \psi'\left(\frac{z+1}{2}\right)- \psi'\left(\frac{z}{2}\right) \right]=-\sum_{r=0}^{\infty} \frac{(-1)^r}{(r+z)^2}
\end{eqnarray}
  is the derivative of the Dirichlet beta function 
  \begin{eqnarray}
\beta(z) =\sum_{r=0}^{\infty} \frac{(-1)^r}{r+z}.
\end{eqnarray}

While the exact physical meaning  of the terms in \eqref{functionsNLO} is still to be clarified, similar functions can be obtained by pole decomposition of powers of the real part   of the digamma fucntion as we show in this study.

The digamma function at natural values of the argument can be related through analytic continuation to the harmonic sum 
\begin{equation}
S_1(n)=\sum_{k=1}^n\frac{1}{k}=\psi(n+1)-\psi(1),
\end{equation}
where $-\psi(1)=\gamma$ is the Euler–Mascheroni constant.

The analytic properties of the digamma function are well studied and known for many years. It has a lot of  functional identities, which are closely related to  the functional identities of $\Gamma(z)$. For example, the reflection formula 
\begin{eqnarray}\label{refl}
\psi(1-z)-\psi(z)=\pi \cot (\pi z)
\end{eqnarray}
can be obatined by differentiating the reflection formula for $\Gamma(z)$.
It follows from ~\eqref{refl} that the point $z_0=\frac{1}{2}$ is the point with respect to which the argument is reflected. In particular, for any complex number  $z=\frac{1}{2}+i b$~($b\in\mathbb{R}$) the expression in ~\eqref{refl} equals twice  the imaginary part of the digamma because $\bar{z}=1-z$ for  $z=\frac{1}{2}+i b$. This implies 
\begin{eqnarray}
\mathfrak{Im} \psi\left(\frac{1}{2}+i b\right)= \frac{\pi}{2}\tanh (\pi b )
\end{eqnarray}
which is readily obtained using the fact that $\cot\left(\frac{\pi}{2}+i \pi b  \right)= \tanh\left( \pi b \right)$. No such closed expression is known for the real part of the digamma function, in part, because there is no reflection formula known for  the sum $\psi(1-z)+\psi(z)$.
However, it turns out that one can write a closed form of the powers of the real part of the digamma function expressed in terms of more complicated functions of either $z=\frac{1}{2}+ i b $ or $\bar{z}=\frac{1}{2}-i b$.
For example, we found that  
\begin{eqnarray}\label{Re2f}
\left(\Re \left[ \psi\left( \frac{1}{2}+i b \right) +\gamma\right]\right)^2&=&\frac{1}{2}  \left( \psi \left(\frac{1}{2}+i b \right)+\gamma \right)^2+\frac{1}{2}  \left( \psi \left(\frac{1}{2}-i b \right)+\gamma \right)^2
\nonumber
\\
&&
-
\frac{1}{4}   \psi^{'} \left(\frac{1}{2}+i b \right)
-
\frac{1}{4}   \psi^{'} \left(\frac{1}{2}-i b \right)+\frac{\pi^2}{4} 
\end{eqnarray}
where $\psi'(z)$ is the derivative of the digamma function.
This calculation is based on the reflection identities for the harmonic sums derived in our previous studies~\cite{prygarin1,prygarin2,prygarin3,prygarin4}.  
The harmonic sums are a useful basis for labelling the space of functions used in the perturbative calculations. The NLO BFKL eigenvalue for zero conformal spin can be presented in terms of harmonic sums as it was shown by Costa, Goncalves and 
Penedones~\cite{Costa}. The harmonic sums were also used in the context of the BFKL equation by Kazkov and Kotikov~\cite{sum1} and Kotikov and Velizhanin~\cite{sum2}.
The use of harmonic sums is mostly convenient because it allows to define the space of functions in an unambiguous way provided the all terms of the underlining expression are of the same known  complexity~\cite{maxtrans}, defined by  transcendentality of corresponding constants that appear as limit of those functions at infinite value of the argument. 
Another approach would be to work in the space of Harmonic Polylogarithms~(HPL), that are  uniquely related to the harmonic sums through the Mellin transform. The advantage of using HPL would be the fact that their functional basis is well defined and their analytic continuation to the complex plane is similar to that of the simple  logarithm function. This approach was implemented for studying helicity amplitudes in the Regge kinematics~\cite{prylip1,prylip2,prylip3,prylip4,prylip5,prylip6,prylip7} 
and the Lipatov effective action~\cite{prylipbond1,prylipbond2}. In this study we calculate   powers of the real part of the digamma function using nested harmonic sums and then express our results in terms of the generalized polygamma functions, similar  to that in \eqref{functionsNLO} using the Mellin transform of HPL. 

In term of the harmonic sums  the expression in \eqref{Re2f} is compactly written as follows
\begin{eqnarray}\label{Re2S}
\left( \Re  \left[S_{1} \left(-\frac{1}{2}+i b \right) \right] \right)^2 &=&   S_{11} \left( -\frac{1}{2}+i b \right) 
+    S_{11} \left( -\frac{1}{2}-i b \right)    \nonumber
\\
&&
-\frac{1}{4}  S_{2} \left( -\frac{1}{2}+i b \right) 
-\frac{1}{4}  S_{2} \left( -\frac{1}{2}+i b \right)
+\frac{\pi^2}{6}.
\end{eqnarray}
 This  can be expressed in a non-linear basis of the harmonic sums  
\begin{eqnarray}\label{Re2Ssimple}
\left( \Re  \left[S_{1} \left(-\frac{1}{2}+i b \right) \right] \right)^2 &=& \frac{1}{4} \left( S_{1} \left( -\frac{1}{2}+i b \right) \right)^2+\frac{1}{4} \left( S_{1} \left( -\frac{1}{2}-i b \right)  \right)^2  \nonumber
\\
&&
+\frac{1}{2}  S_{2} \left( -\frac{1}{2}+i b \right) 
+\frac{1}{2}  S_{2} \left( -\frac{1}{2}+i b \right)
+\frac{\pi^2}{6}.
\end{eqnarray}

Next we consider the real part of  $S_1\left(-\frac{1}{2}+ib\right)$, i.e. the real part of $\psi \left(-\frac{1}{2}+ib\right)-\psi(1)$, to the third  power, namely, 
 $\left(\Re \left[ S_1\left(-\frac{1}{2}+ib\right) \right]\right)^3$. It can be obtained  multiplying  the expression in  ~\eqref{Re2S} by the real part of $S_1(z)$
 \begin{eqnarray}
 \Re \left[S_1\left(-\frac{1}{2}+i b\right)\right]\equiv\frac{1}{2 } \left(  S_1\left(-\frac{1}{2}+i b\right)+S_1\left(-\frac{1}{2}-i b\right)\right).
\end{eqnarray}  
That multiplication includes    crossed terms of two types  $S_{1}(z)S_{11}(\bar{z})$ and $S_{1}(z)S_{2}(\bar{z})$, where $z=-\frac{1}{2}+i b$ and $\bar{z}=-\frac{1}{2}-i b$.
 The cross terms can be further written using pole decomposition obtained with the help of the  reflection identities for harmonic sums calculated in \cite{Prygarin:2018tng} for weight three (for weight  four see \cite{Prygarin:2018cog}), namely 
 \begin{eqnarray}\label{refls1sc2}
&& S_1(z) S_2(-1-z)= 3 \zeta (3)-\frac{1}{6} \pi ^2 S_1(-1-z)+\frac{1}{6} \pi ^2
   S_1(z)
 +S_{1,2}(-1-z)-S_{2,1}(z)
\end{eqnarray}
 and 
\begin{eqnarray}\label{refls1sc11}
&& S_1(z) S_{1,1}(-1-z)= 3 \zeta (3)+\frac{1}{6} \pi ^2 S_1(-1-z)+\frac{1}{6} \pi ^2
   S_1(z)-S_{2,1}(-1-z)
   \nonumber
    \\
 && 
 \hspace{1.9cm} 
 -S_{2,1}(z)+2 S_{1,1,1}(-1-z)+S_{1,1,1}(z),
\end{eqnarray}
 where $z=-\frac{1}{2}+i b$ and  $\bar{z}=-1-z=-\frac{1}{2}-i b$.  
    
The final expression reads    
\begin{eqnarray}
\left( \Re  \left[S_{1} \left(-\frac{1}{2}+i b \right) \right] \right)^3 &=& \frac{1}{4} \pi ^2 S_1(z)+\frac{1}{4} \pi ^2 S_1(\bar{z})+
\frac{1}{2} (S_1(z))^3 +\frac{1}{2} (S_1(\bar{z}))^3 
\nonumber
\\
&&
+\frac{3}{8} S_3(z)+\frac{3}{8} S_3(\bar{z})-\frac{3}{4} S_{2,1}(z)-\frac{3}{4} S_{2,1}(\bar{z}) \nonumber
\\
&&
+\frac{3}{4} S_2(z) S_1(z)+\frac{3}{4} S_1(\bar{z}) S_2(\bar{z})+\frac{9 \zeta (3)}{4},
\end{eqnarray}  
 which can be written in a linear basis as follows

\begin{eqnarray}
\left( \Re  \left[S_{1} \left(-\frac{1}{2}+i b \right) \right] \right)^3 &=& 
-\frac{3}{4} S_{1,2}(z)-\frac{3}{2} S_{2,1}(z)
+3 S_{1,1,1}(z)-\frac{3}{4}
   S_{1,2}(\bar{z})-\frac{3}{2} S_{2,1}(\bar{z})
   \\
&&
+3
   S_{1,1,1}(\bar{z})+\frac{1}{4} \pi ^2
   S_1(z)
  +\frac{S_3(z)}{8}+\frac{1}{4} \pi ^2
   S_1(\bar{z})+\frac{S_3(\bar{z})}{8}+\frac{9 \zeta (3)}{4}.
   \nonumber
\end{eqnarray}

This expression is symmetrical with respect to $z \leftrightarrow \bar{z}$ and can be compactly written
 
\begin{eqnarray}\label{Re3}
\left( \Re  \left[S_{1} \left(-\frac{1}{2}+i b \right) \right] \right)^3 = F_3\left(-\frac{1}{2}+i b\right)+F_3\left(-\frac{1}{2}-i b\right)
\end{eqnarray}
 in terms 
of the function $F_3(z)$ given by 
\begin{eqnarray}
F^{lin}_3(z)=-\frac{3}{4} S_{1,2}(z)-\frac{3}{2} S_{2,1}(z)+3 S_{1,1,1}(z)+\frac{1}{4}
   \pi ^2 S_1(z)+\frac{S_3(z)}{8}+\frac{9 \zeta (3)}{8}
\end{eqnarray}
or  in the non-linear basis 
\begin{eqnarray}\label{F3nonlin}
F^{non-lin}_3(z)=-\frac{3}{4} S_{2,1}(z)+\frac{1}{2} S_1(z){}^3+\frac{3}{4} S_2(z)
   S_1(z)+\frac{1}{4} \pi ^2 S_1(z)+\frac{3 S_3(z)}{8}+\frac{9 \zeta (3)}{8}.
\end{eqnarray}  
 Those two representations are completely equivalent due to the quasi-shuffle algebra of the harmonic sums,  and the choice of the linear or non-linear basis is merely a matter of convenience. 

 The harmonic sums require analytic continuation to the complex plane. For harmonic sums of one positive index the analytic continuation is rather simple
\begin{eqnarray}\label{saf}
S_a (z) \Rightarrow - \sum_{j=0}^{\infty} \frac{1}{(j+z+1)^a} +\zeta(a)=\zeta (a)-\frac{(-1)^a \psi ^{(a-1)}(z+1)}{(a-1)!},
\;\; a>1
\end{eqnarray} 
 where $\psi^{(a)}(z)$ denotes the derivative of the polygamma function of order $a$. We skip the discussion of the analytic continuation in general case and quote the only harmonic sum of two indices  $S_{21}(z)$ appearing in ~\eqref{F3nonlin}
\begin{eqnarray}\label{s21f}
S_{21}(z) \Rightarrow 2 \zeta (3)-\sum _{j=0}^{\infty } \frac{\psi'(j+1)}{j+z+1}=\int_0^1 x^z \frac{\text{Li}_2(x) -\frac{\pi ^2}{6} }{ 1-x} dx +2 \zeta (3)
 \end{eqnarray}
 where $\zeta(a)$ stands for the Riemann zeta function and  $\psi'(z)$ is the trigamma function, i.e the first derivative of the digamma function. The analytic continuation of the harmonic sum $S_{21}(z)$ to the complex plane can be done in either series or integral representation, as shown in ~\eqref{s21f}, and the choice of the representation is a matter of convenience.

Finally we write  the third power of the real part of  digamma function given in ~\eqref{Re3} in the analytic functional form 
\begin{eqnarray}\label{Re3f}
\left(\Re \left[ \psi\left(\frac{1}{2}+i b  \right)-\psi(1)\right]\right)^3=F_3\left(-\frac{1}{2}+i b  \right)+F_3\left(-\frac{1}{2}-i b  \right),
\end{eqnarray} 
 where $F_3(z)$ reads 
\begin{eqnarray}\label{F3fI}
F_3(z)&=&\frac{3 }{4}
\sum _{j=0}^{\infty } \frac{\psi'(j+1)}{j+z+1}
+\frac{\psi  (z+1)^3}{2}+\frac{3}{2} \gamma  \psi
    (z+1)^2-\frac{3}{4} \psi'(z+1) \psi  (z+1) \nonumber
    \\
    &&+\frac{3}{8}
   \pi ^2 \psi  (z+1)+\frac{3}{2} \gamma ^2 \psi
  (z+1)-\frac{3}{4} \gamma  \psi '(z+1)\nonumber
  \\
    &&+\frac{3 \psi'' (z+1)}{16}+\frac{3 \gamma  \pi ^2}{8}+\frac{\gamma ^3}{2}, 
\end{eqnarray} 
 The sum of the first term can be replaced by the corresponding integral 
 \begin{eqnarray}
 \sum _{j=0}^{\infty } \frac{\psi'(j+1)}{j+z+1}=-\int_0^1  x^z \frac{\text{Li}_2(x) -\frac{\pi ^2}{6}}{ 1-x } dx
 \end{eqnarray}
 in accordance with the analytic continuation of $S_{21}(z)$ in  ~\eqref{s21f}. The integral representation is more efficient for numerical calculations, whereas the series representation shows explicitly the pole structure of $S_{21}(z)$.
 \newpage

 \section{Fourth power of the real part of the digamma function}
In this section we proceed with  calculating the fourth power of the  real part of the digamma function multiplying the third  power of the real part of $S_1(z)$ in ~\eqref{Re3} by the expression in ~\eqref{Re2Ssimple}. We use the reflection identities for the harmonic sums rewriting cross terms of the form $S_{a_1,a_2,...}(z) S_{b_1,b_2,...}(1-z)$ as a sum of holomorphic and antiholomorphic parts.
 Details of calculations are presented below and the final result can be written as follows
\begin{eqnarray}\label{Re4}
\left( \Re  \left[S_{1} \left(-\frac{1}{2}+i b \right) \right] \right)^4 = F_4\left(-\frac{1}{2}+i b\right)+F_4\left(-\frac{1}{2}-i b\right)
\end{eqnarray}
 in terms 
of function $F_4(z)$ given by 
\begin{eqnarray}
F^{lin}_4(z)&=& \pi ^2 S_{1,1}(z)+\frac{1}{2} S_{1,3}(z)+\frac{3}{2} S_{2,2}(z)+2
   S_{3,1}(z)-3 S_{1,1,2}(z)-6 S_{1,2,1}(z) \nonumber
   \\
   &&-6 S_{2,1,1}(z)+12
   S_{1,1,1,1}(z)+\frac{9}{2} \zeta (3) S_1(z)-\frac{1}{3} \pi ^2
   S_2(z)-\frac{S_4(z)}{16}+\frac{53 \pi ^4}{720}
\end{eqnarray}
This can be expressed in the non-linear basis using the quasi-shuffle algebra of the harmonic sums
\begin{eqnarray}\label{F4nonlin}
F^{non-lin}_4(z)&=&-3 S_1(z) S_{2,1}(z)-\frac{3}{2} S_{3,1}(z)+3 S_{2,1,1}(z)+\frac{9}{2} \zeta
   (3) S_1(z)+\frac{1}{2} S_1(z){}^4\nonumber
   \\
   &&+\frac{3}{2} S_2(z)
   S_1(z){}^2+\frac{1}{2} \pi ^2 S_1(z){}^2+\frac{3}{2} S_3(z)
   S_1(z)-\frac{3}{4} S_2(z){}^2\nonumber
   \\
   &&+\frac{1}{6} \pi ^2 S_2(z)
   -\frac{5
   S_4(z)}{16}+\frac{53 \pi ^4}{720}
\end{eqnarray}
 
The harmonic sums  appearing in~\eqref{F4nonlin} require   analytic continuation and most of them were discussed in the previous section. The harmonic sums that  here and were discussed in the previous section  are $S_{3,1}(z)$, $S_{2,2}(z)$ and $S_{2,1,1}(z)$.  The  analytic continuation  of $S_{3,1}(z)$ is given by
\begin{eqnarray}\label{s31f}
S_{31}(z) &\Rightarrow&
-\frac{1}{2}\sum_{j=0}^{\infty}\frac{\psi''(j+1)}{j+z+1}-\frac{\pi^2}{6}  \psi'(z+1)+\frac{\pi ^4}{72} \nonumber 
   \\
   &=&
\frac{\pi ^4}{72}
-\frac{\pi^2}{6}  \psi'(z+1)-\int_0^1 x^z \frac{ \text{Li}_3(x)-\zeta(3)}{1-x} dx.
\end{eqnarray} 

The harmonic sums $S_{2,2}(z)$ and $S_{2,1,1}(z)$ can be expressed as Mellin transform of Harmonic Polylogarithms as follows. The Mellin transform which stands for $S_{2,2}(z)$ reads
\begin{eqnarray}\label{s22f}
S_{2,2}(z) \Rightarrow  
-\int_{0}^1 x^z \frac{H_{0,1,0}(x)+2 \zeta(3)}{1-x}
+\frac{7 \pi ^4}{360},
\end{eqnarray}
where 
  \begin{eqnarray}\label{H010}
H_{0,1,0}(x)= \ln (x)\text{Li}_2(x)  -2\text{Li}_3(x)
 \end{eqnarray}

The integral representation of $S_{2,1,1}(z)$ is given by 
\begin{eqnarray}\label{s211f}
S_{2,1,1}(z) &\Rightarrow&   \int_0^1  x^z \frac{H_{0,1,1}(x)-\zeta(3)}{1-x}
+\frac{ \pi ^4}{30},
\end{eqnarray}
where
   \begin{eqnarray}\label{H011}
H_{0,1,1}(x)= -\text{Li}_3(1-x)+\text{Li}_2(1-x) \log (1-x)+\frac{1}{2} \log (x) \log
   ^2(1-x)+\zeta (3).
 \end{eqnarray}

 It worth mentioning that  the integrals in \eqref{s21f},\eqref{s31f}, \eqref{s22f} and \eqref{s211f} are convergent ones.
  
 Using the integral representation of the analytic continuation of the harmonic sums in \eqref{saf},\eqref{s21f}, \eqref{s31f}, \eqref{s211f} express the fourth power  of the real part  in \eqref{Re4} 
\begin{eqnarray}\label{Re4f}
\left( \Re  \left[\psi \left(\frac{1}{2}+i b \right)-\psi(1) \right] \right)^4 = F_4\left(-\frac{1}{2}+i b\right)+F_4\left(-\frac{1}{2}-i b\right)
\end{eqnarray}
in the analytic form    
\begin{eqnarray}\label{F4fI}
\hspace{-1cm}F_4(z)&=&-3 \gamma  \mathtt{I}_1(z)-3 \mathtt{I}_1(z) \psi (z+1)+\frac{3
   \mathtt{I}_2(z)}{2}+3 \mathtt{I}_4(z)+\frac{\psi (z+1)^4}{2}+2 \gamma 
   \psi (z+1)^3 \nonumber \\
   &&-\frac{3}{2} \psi '(z+1) \psi
   (z+1)^2+\frac{3}{4} \pi ^2 \psi (z+1)^2+3 \gamma ^2 \psi
   (z+1)^2-3 \gamma  \psi '(z+1) \psi (z+1)\nonumber 
   \\
   &&+\frac{3}{4}
   \psi ^{''}(z+1) \psi (z+1)+\frac{3}{2} \gamma  \pi ^2 \psi
   (z+1)+2 \gamma ^3 \psi (z+1)-\frac{3 \psi
   '(z+1)^2}{4} +\frac{\gamma ^4}{2} \nonumber 
   \\
   &&
   +\frac{1}{3} \pi ^2 \psi '(z+1)-\frac{3}{2} \gamma
   ^2 \psi '(z+1)+\frac{3}{4} \gamma  \psi ^{''}(z+1)+\frac{5 \psi
   ^{'''}(z+1)}{96}+\frac{5 \pi ^4}{32}+\frac{3 \gamma ^2 \pi
   ^2}{4},
\end{eqnarray}
where $\mathtt{I}_i(z)$  read
\begin{eqnarray}\label{I1}
\mathtt{I}_1(z)=\int_0^1 x^z\frac{  \texttt{Li}_2(x) -\frac{\pi^2}{6}}{1-x}dx,
\end{eqnarray} 

\begin{eqnarray}\label{I2}
\mathtt{I}_2(z)= \int_0^1 x^z \frac{\mathtt{Li}_3(x)-\zeta(3)}{1-x} dx 
\end{eqnarray}

\begin{eqnarray}\label{I3}
\mathtt{I}_3(z)= \int_0^1 x^z \frac{\mathtt{H}_{0,1,0}(x)+2 \zeta(3)}{1-x} dx 
\end{eqnarray}

\begin{eqnarray}\label{I4}
\mathtt{I}_4(z)= \int_0^1 x^z \frac{\mathtt{H}_{0,1,1}(x)- \zeta(3)}{1-x} dx 
\end{eqnarray}
 with the harmonic polylogariths  $\mathtt{H}_{0,1,0}(x)$ and $\mathtt{H}_{0,1,1}(x)$     given by \eqref{H010} and \eqref{H011}. Note that  the expressions in $\mathtt{I}_i(z)$ are finite, which is related to the fact that the harmonic polylogariths appearing in $\mathtt{I}_i(z)$ are approaching unity as  $\mathtt{H}_{a,b,c,..}(x)-\mathtt{H}_{a,b,c,..}(1) \simeq \mathcal{O}(x^n), \;x\geq 1$. The specific values of  harmonic polylogariths in $\mathtt{I}_i(z)$ at unity  are given by 
 \begin{eqnarray}
 \mathtt{Li}_2(1)=\frac{\pi^2}{6}, \;  \mathtt{Li}_3(1)=\zeta(3), \; \mathtt{H}_{0,1,0}(1)=-2 \zeta(3), \;  \mathtt{H}_{0,1,1}(1)=\zeta(3).
 \end{eqnarray}

\section{Pole Structure} 
 
 In this section we analyze the analytic structure of the functions, which build expressions in \eqref{F3fI} and \eqref{F4fI}. 
 Those expressions are built of polygamma functions 
 \begin{eqnarray}
 \psi^{(a)} (z)=\frac{d^a \psi(z)}{d z^a } = (-1)^{a+1} a! \sum_{j=0}^{\infty} \frac{1}{(j+z+1)^{a+1}}, \;\; a \geq 1
\end{eqnarray}   
 and their generalizations defined in \eqref{I1}, \eqref{I2}, \eqref{I3} and \eqref{I4}. The generalized polygamma functions can be written in   series representation, which makes their analytic explicit. For example, consider $\mathtt{I}_1(z)$  in \eqref{I1}  and its series representation 
 \begin{eqnarray}
 \mathtt{I}_1(z)= \int_{0}^1  x^z \frac{\mathtt{Li}_2(x)-\frac{\pi^2}{6}}{1-x} dx =-\sum_{j=0}^{\infty} \frac{\psi'(j+1)}{j+z+1}   
  \end{eqnarray}
 which reveals isolated poles at negative integers. It is worth emphasizing that those poles are of the same order and    equidistant, but have different residues. The residues at adjacent poles $z_k$ and $z_{k+1}$  are related by a shift of $1/(z_k+1)^2$ due the recurrence relation~("telescope identity") of the trigamma function. 
 \begin{figure}[h]
\centering
\includegraphics[scale=0.65]{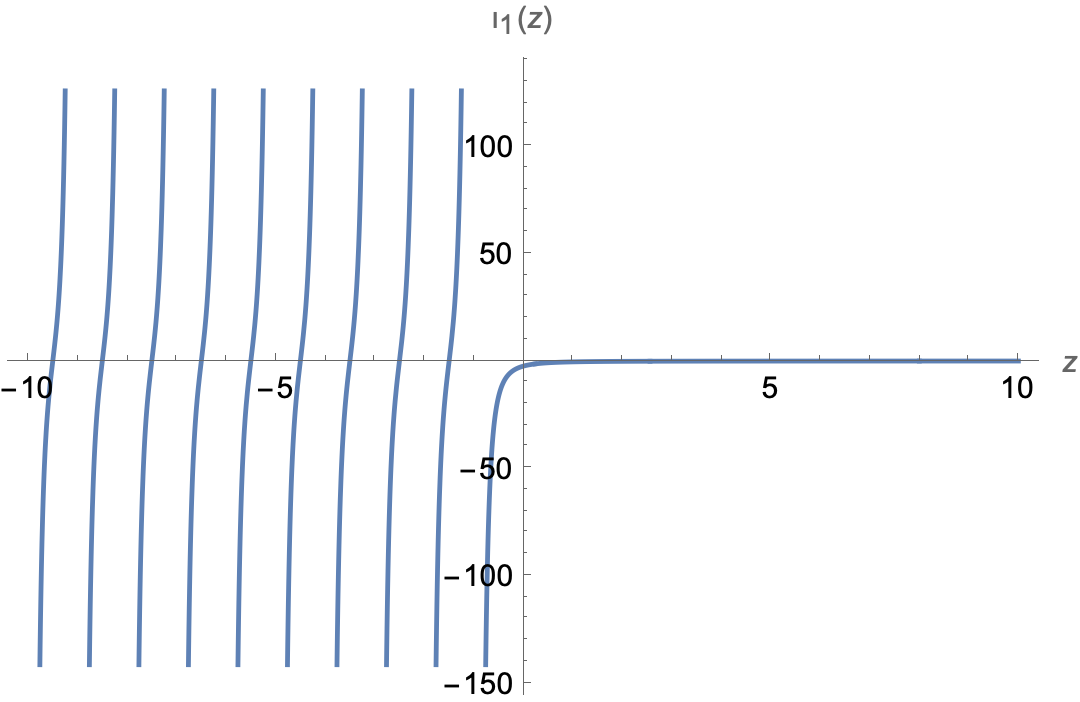}\caption{The figure depicts the plot of  $\mathtt{I}_1(z)$, which has isolated simple poles at negative integer values of $z$. This pole structure is similar to the pole structure of even derivatives of the digamma function $\psi^{(2k)}(z+1), \;\; k=1,2,3,...$.   }\label{fig:polesodd1}
\end{figure}
 
 Fig.~\ref{fig:polesodd1} illustrates  the pole stricture of 
  $\mathtt{I}_1(z)$, which has simple isolated poles at negative integer values of its argument. 
 A similar pole structure has 
  \begin{eqnarray}\label{I2Sum}
 \mathtt{I}_2(z) = \int_{0}^1  x^z \frac{\mathtt{Li}_3(x)-\zeta(3)}{1-x} dx =\frac{1}{2}\sum_{j=0}^{\infty} \frac{\psi''(j+1)}{j+z+1} 
  \end{eqnarray}
  and

\begin{eqnarray}\label{I3Sum}
\mathtt{I}_3(z)= \int_0^1 x^z \frac{\mathtt{H}_{0,1,0}(x)+2 \zeta(3)}{1-x} dx = \sum_{j=0}^{\infty} \frac{\psi'(j+1)}{(j+z+1)^2}-\frac{\pi^2}{6}\psi'(z+1)-2 \mathtt{I}_2(z).
\end{eqnarray}
    A similar series representation can be found for $\mathtt{I}_4(z)$  in \eqref{I4}, which is beyond the scope of the present study. The similarities in the analytic structure of all $\mathtt{I}_i(z)$ can be inspected by noting that  their integrand is constant at the upper limit as  $x \to 1$ and diverges at the lower limit $x \to 0 $  only at $z=-k, \; k \in \mathbb{N}$. 
  The series representation is very useful in analysing the analytic structure of the functions appearing in   \eqref{F3fI} and   \eqref{F4fI}, which have poles at negative integer values of the argument. The maximal pole order corresponds to the power of the real part of the digamma function. 
 
 The expressions in \eqref{Re2f}, \eqref{Re3f} and \eqref{Re4f} of powers of the real part of the digamma function 
 $\left( \Re  \left[\psi \left(\frac{1}{2}+i b \right)-\psi(1) \right] \right)^k$ provide a useful separation of the poles on the right and left part of the complex plane and present the main result of this study. 
  
 \section{Conclusions}
 In this paper we consider powers of the real part of the digamma function as a functions of complex variable $z=\frac{1}{2}+ib,\;\; b \in \mathtt{Reals}$. We decompose it into two symmetric pieces, which have  symmetrical poles located on either side of the line $\Re ( z) =\frac{1}{2}$.
The pole decomposition of  third and the fourth power of the real part of the digamma function in  \eqref{Re3f} and  \eqref{Re4f}, together with their explicit form in \eqref{F3fI} and \eqref{F4fI} present the main result of the present study. 
 This  result has a  transparent pole structure, but it includes generalized polygamma functions, which are much more complicated than the digamma function.    
 
  A similar structure appears exponentiating the BFKL eigenvalue at the leading order for zero conformal spin. We show that exponentiating a relatively simple digamma function generates complicated generalizations of polygamma function, which appear in the BFKL eigenvalue at higher orders of the perturbation theory.

\section{Acknowledgements} 
We are grateful to Sergey Bondarenko for fruitful discussions on main topics covered here. This work is supported in part by "Program of Support of High Energy Physics" Grant by Israeli Council for Higher Education.

\end{document}